\documentclass{llncs}
\usepackage[show]{ed}
\usepackage[utf8]{inputenc}
\usepackage{lststex}
\usepackage{paralist}
\usepackage{tikz,standalone}
\usetikzlibrary{backgrounds,shapes,fit,shadows,mmt}
\usepackage{wrapfig}
\usepackage{xspace}
\usepackage{hyperref}
\usepackage{stex-logo}
\usepackage{odksystems}
\usepackage[style=alphabetic,backend=bibtex]{biblatex}
\renewbibmacro*{event+venue+date}{}
\renewbibmacro*{doi+eprint+url}{%
  \iftoggle{bbx:doi}
    {\printfield{doi}\iffieldundef{doi}{}{\clearfield{url}}}
    {}%
  \newunit\newblock
  \iftoggle{bbx:eprint}
    {\usebibmacro{eprint}}
    {}%
  \newunit\newblock
  \iftoggle{bbx:url}
    {\usebibmacro{url+urldate}}
    {}}
\title{Interoperability in the \ODK Project:\\
The Math-in-the-Middle Approach}
\author{Paul-Olivier Dehaye\inst{3} Michael Kohlhase\inst{2} Alexander
  Konovalov\inst{4} Samuel Lelièvre\inst{1} Markus
  Pfeiffer\inst{4} Nicolas M. Thiéry\inst{1}}
\institute{
  Universit\'e Paris-Sud\and
Jacobs University \and 
  University of Z\"urich \and
  University of St~Andrews
}
\providecommand\ifprefchar[2]{}%
\newcommand\defemph[1]{\textbf{#1}}
\renewcommand{\ednote}[1]{}

\begin{document}
\maketitle
\begin{abstract}
  \ODK --- ``Open Digital Research Environment Tool\-kit for the Advancement of
  Mathematics'' --- is an H2020 EU Research Infrastructure project that aims at
  supporting, over the period 2015--2019, the ecosystem of open-source mathematical
  software systems. From that, \ODK will deliver a flexible toolkit enabling research
  groups to set up Virtual Research Environments, customised to meet the varied needs of
  research projects in pure mathematics and applications.

  An important step in the \ODK endeavor is to foster the interoperability between a
  variety of systems, ranging from computer algebra systems over mathematical databases to
  front-ends. This is the mission of the integration work package (WP6). We report on
  experiments and future plans with the \emph{Math-in-the-Middle} approach. This
  information architecture consists in a central mathematical ontology that documents the
  domain and fixes a joint vocabulary, combined with specifications of the functionalities
  of the various systems. Interaction between systems can then be enriched by pivoting off
  this information architecture.
\end{abstract}

\section{Introduction}

From their earliest days, computers have been used in pure mathematics, either to make
tables, to prove theorems (famously the four colour theorem) or, as with the astronomer's
telescope, to explore new theories. Computer-aided experiments, and the use of databases
relying on computer calculations such as the Small Groups Library in GAP, the Modular
Atlas in group and representation theory, or the $L$-functions and Modular Forms Database (\LMFDB, see later), are part of the standard
toolbox of the pure mathematician, and certain areas of mathematics completely depend on
it. Computers are also increasingly used to support collaborative work and education.

The last decades witnessed the emergence of a wide ecosystem of open-source tools to
support research in pure mathematics. This ranges from specialized to general purpose
computational tools such as \GAP, \PariGP, \Linbox, \MPIR, \Sage, or \Singular, via online
databases like the \LMFDB and does not count online services like Wikipedia,
\Arxiv, or MathOverflow. A great opportunity is the rapid emergence of key technologies,
and in particular the \Jupyter (previously \IPython) platform for interactive and
exploratory computing which targets all areas of science.

This has proven the viability and power of collaborative open-source development models,
by users and for users, even for delivering general purpose systems targeting a large
public (researchers, teachers, engineers, amateurs, \ldots). Yet some critical long term
investments, in particular on the technical side, are in order to boost the productivity
and lower the entry barrier:
\begin{compactitem}
\item Streamlining access, distribution, portability on a wide range of platforms, including
  High Performance Computers or cloud services.
\item Improving user interfaces, in particular in the promising area of collaborative
  workspaces as those provided by \SMC.
\item Lowering barriers between research communities and promote dissemination. For example
  make it easy for a specialist of scientific computing to use tools from pure
  mathematics, and reciprocally.
\item Bringing together the developers communities to promote tighter collaboration and
  symbiosis, accelerate joint development, and share best practices.
\item Outsourcing as much of the development as possible to larger communities to focus
  the work forces on their core specialty: the implementation of mathematical algorithms
  and databases.
\item And last but not least: Promoting collaborations at all scales to further improve
  the productivity of researchers in pure mathematics and applications.
\end{compactitem}
These can be subsumed by the goal of \emph{Virtual Research Environments} (VRE), that is
online services enabling groups of researchers, typically widely dispersed, to work
collaboratively on a per project basis. This is exactly where the \ODK project kicks in. 

We will introduce the \ODK project Section~\ref{sec:odk} to establish the context for the
``Math-in-the-Middle'' (MitM) integration approach described in
Section~\ref{sec:mitm}. The remaining sections then elucidate the approach by presenting
first experiments and refinements of the chosen integration paradigm:
Section~\ref{sec:lmfdb} details how existing knowledge representation and data structures
can be represented as MitM interface theories with a case study of equipping the \LMFDB
with a MitM-based programming interface.  Section~\ref{sec:gapsage} discusses system
integration between \GAP and \Sage and how this can be routed through a MitM
ontology. Section~\ref{sec:concl} concludes the paper and discusses future work.

\section{The \ODK Project}\label{sec:odk}
The project ``Open Digital Research Environment Toolkit for the Advancement of
Mathematics'''~\cite{OpenDreamKit:on} is a European H2020 project funded under the
EINFRA-9 call~\cite{EINFRA-9} whose theme of \emph{Virtual Research Environments} was a
natural fit to seek for manpower and funding for a developer community that have been
working really hard on the items above. The \ODK consortium consists of core European
developers of the aforementioned systems for pure mathematics, and reaching toward the
numerical community, and in particular the \Jupyter community, to work together on joint
needs. The project aims to address the following goals in close collaboration with the
community:
\begin{compactenum}
\item Further improve the productivity of researchers in pure mathematics and applications
  by further promoting collaborations on \emph{Data}, \emph{Knowledge}, and
  \emph{Software}.
\item Make it easy for teams of researchers of any size to set up custom, collaborative
  \emph{Virtual Research Environments} tailored to their specific needs, resources, and
  workflows.
\item Support the entire life-cycle of computational work in mathematical research, from
  \emph{initial exploration} to \emph{publication}, \emph{teaching}, and \emph{outreach}.
\end{compactenum}
The acceptance of the proposal~\cite{ODKproposal:on} in May 2015 was a strong sign of
recognition, at the highest level of funding agencies, of the values of open science and
the strength and maturity of the ecosystem.

The \ODK projects~\cite{ODKproposal:on} will run for four years, starting from September
2015. It involves about 50 people spread over 15 sites in Europe, with a total budget of
about 7.6 million euros. The largest portion of that will be devoted to employing an
average of 11 researchers and developers working full time on the project. Additionally,
the participants will contribute the equivalent of six other people working full time.  By
definition this project will be mostly funding actions in Europe; however those actions
will be carried out, as usual, in close collaborations with the worldwide community
(potential users of the VRE as well as developers outside the \ODK consortium).

The \ODK work plan consists in 58 concrete tasks split in seven work packages, which
include:
\begin{description}
\item[WP3: Component Architecture] work on portability -- especially
  on the Windows platform -- modularity, packaging, distribution,
  deployment, standardization and interoperability between components.
\item[WP4: User Interfaces] work on uniform \Jupyter notebook
  interfaces for all interactive computational components, improvements
  to \Jupyter, 3D visualization, documentation tools, ...
\item[WP5: High Performance Mathematical Computing] work within
  and between ODK's components to improve performance, and in
  particular better exploit multicore / parallel architectures.
\item[WP6: Data/Knowledge/Software-Bases] \emph{e.g.}\ identification and extensions of ontologies
  and standards to facilitate safe and efficient storage, reuse, interoperation and
  sharing of rich mathematical data, whilst taking into account of provenance and
  citability; data archiving and sharing in a semantically sound way component
  architecture; integration between computational software and databases.
\item[WP7: Social Aspects] research on social aspects of collaborative
  Data, Knowledge, Software development in mathematics to inform the
  other WPs.
\end{description}
\ODK will also actively engage in community building and training by organizing workshops
and training materials.

An innovative aspect of the \ODK project is that its preparation and management happens,
as much as is practical and without infringing on privacy, in the open. For example, most
documents, including the proposal itself, are version controlled on public repositories
and progress on tasks and deliverables is tracked using public issues
(see~\cite{OpenDreamKit:on}). This has proven a strong feature to collaborate tightly with
the community and get early feedback.

\section{Integrating Mathematical Software Systems via the Math-in-the-Middle Approach}\label{sec:mitm}

To achieve the goal of assembling the ecosystem of mathematical software systems in the
\ODK project into a coherent mathematical VRE, we have to make the systems interoperable at
a mathematical level. In particular, we have to establish a common meaning space that
allows to share computation, visualization of the mathematical concepts, objects, and
models (COMs) between the respective systems. Building on this we can build a VRE with
classical techniques for integrated development environments (IDE).

\subsection{A Common Meaning Space for Interoperability}

Concretely, the problem is that the software systems in \ODK have different coverage, and
where these overlap representations of and functionalities for the COMs involved differ.
This starts with simple naming issues (\emph{e.g.}\ elliptic curves are named
\lstinline|ec| in the \LMFDB, and as \lstinline|EllipticCurve| in \Sage), persists through
the underlying data structures (permutations are represented as products of cycles in
\GAP, in list form in \Sage, and in differing representations in the various tables of the
\LMFDB), and becomes virulent at the level of algorithms, their parameters, and domains of
applicability.

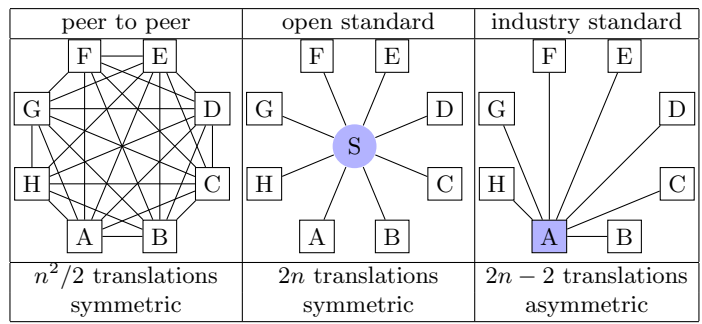
\begin{figure}[ht]\centering
  \begin{tabular}{|c|c|c|}\hline
    peer to peer & open standard & industry standard\\\hline
        \begin{tikzpicture}
      \node[draw] (a) at (0,.3) {A};
      \node[draw] (b) at (1,.3) {B};
      \node[draw] (c) at (1.7,1) {C};
      \node[draw] (d) at (1.7,2) {D};
      \node[draw] (e) at (1,2.7) {E};
      \node[draw] (f) at (0,2.7) {F};
      \node[draw] (g) at (-.7,2) {G};
      \node[draw] (h) at (-.7,1) {H};
      \draw (a) -- (b) -- (c) -- (d) -- (e) -- (f) -- (g) -- (h) -- (a);
      \draw (a) -- (c) -- (h) -- (d) -- (g) -- (e);
      \draw (b) -- (h);
      \draw (b) -- (f);
      \draw (b) -- (d);
      \draw (b) -- (g);
      \draw (b) -- (e);
      \draw (h) -- (e);
      \draw (d) -- (f);
      \draw (g) -- (c);
      \draw (a) -- (d);
      \draw (a) -- (e);
      \draw (a) -- (f);
      \draw (a) -- (g);
      \draw (e) -- (c);
      \draw (c) -- (f);
    \end{tikzpicture} &     \begin{tikzpicture}
      \node[draw] (a) at (0,.3) {A};
      \node[draw] (b) at (1,.3) {B};
      \node[draw] (c) at (1.7,1) {C};
      \node[draw] (d) at (1.7,2) {D};
      \node[draw] (e) at (1,2.7) {E};
      \node[draw] (f) at (0,2.7) {F};
      \node[draw] (g) at (-.7,2) {G};
      \node[draw] (h) at (-.7,1) {H};
      \node[circle,fill=blue!30] (m) at (.5,1.5) {S};
      \draw (m) -- (a);
      \draw (m) -- (b);
      \draw (m) -- (c);
      \draw (m) -- (d);
      \draw (m) -- (e);
      \draw (m) -- (f);
      \draw (m) -- (g);
      \draw (m) -- (h);
    \end{tikzpicture} &     \begin{tikzpicture}
      \node[draw,fill=blue!30] (a) at (0,.3) {A};
      \node[draw] (b) at (1,.3) {B};
      \node[draw] (c) at (1.7,1) {C};
      \node[draw] (d) at (1.7,2) {D};
      \node[draw] (e) at (1,2.7) {E};
      \node[draw] (f) at (0,2.7) {F};
      \node[draw] (g) at (-.7,2) {G};
      \node[draw] (h) at (-.7,1) {H};
      \draw (a) -- (b);
      \draw (a) -- (c);
      \draw (a) -- (d);
      \draw (a) -- (e);
      \draw (a) -- (f);
      \draw (a) -- (g);
      \draw (a) -- (h);
    \end{tikzpicture}\\\hline
    $n^2/2$  translations & $2n$ translations & $2n-2$ translations \\
    symmetric & symmetric & asymmetric\\\hline
  \end{tabular}
  \caption{Approaches for many-systems interoperability}\label{fig:interop}
\end{figure}

To obtain a common meaning space for a VRE, we have the three well-known approaches in
Figure~\ref{fig:interop}.

The first does not scale to a project with about a dozen systems, for the third there is
no obvious contender in the \ODK ecosystem. Fortunately, we already have a ``standard'' for
expressing the meaning of COMs -- \defemph{mathematical vernacular}: the language of
mathematical communication, and in fact all the COMs supported in the \ODK VRE are documented
in mathematical vernacular in journal articles, manuals, etc.

The obvious problem is that mathematical vernacular is too 
\begin{inparaenum}[\em i\rm)]
\item \emph{ambiguous}: we need a human to understand structure, words, and symbols
\item \emph{redundant}: every paper introduces slightly different notions. 
\end{inparaenum}

Therefore we explore an approach, where we partially formalize (\defemph{flexiformalize};
see~\cite{Kohlhase:tffm13}) mathematical vernacular to obtain a flexiformal ontology of
mathematics that can serve as an open communication vocabulary. We call the approach the
\defemph{Math-in-the-Middle} (MitM) Strategy for integration and the ontology the \defemph{MitM
ontology}.

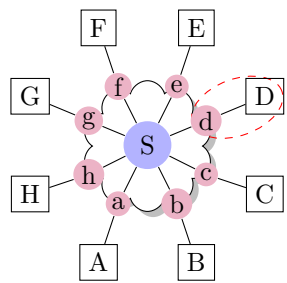
\begin{wrapfigure}r{4cm}\vspace*{-1.5em}
      \begin{tikzpicture}[scale=1.3]
      \tikzstyle{withshadow}=[draw,drop shadow={opacity=.5},fill=white]
      \tikzstyle{system}=[draw]
      \tikzstyle{standard}=[circle,fill=blue!30]
      \tikzstyle{interface}=[circle,fill=purple!30,inner sep = 1pt,]
      \node[system] (a) at (0,.3) {A};
      \node[system] (b) at (1,.3) {B};
      \node[system] (c) at (1.7,1) {C};
      \node[system] (d) at (1.7,2) {D};
      \node[system] (e) at (1,2.7) {E};
      \node[system] (f) at (0,2.7) {F};
      \node[system] (g) at (-.7,2) {G};
      \node[system] (h) at (-.7,1) {H};
      \node[standard] (m) at (.5,1.5) {S};
      \node[interface] (ia) at (0.2,.9) {a};
      \node[interface] (ib) at (.8,.9) {b};
      \node[interface] (ic) at (1.1,1.2) {c};
      \node[interface] (id) at (1.1,1.75) {d};
      \node[interface] (ie) at (.8,2.1) {e};
      \node[interface] (if) at (0.2,2.1) {f};
      \node[interface] (ig) at (-.1,1.75) {g};
      \node[interface] (ih) at (-.1,1.2) {h};
      \draw (m) -- (ia) -- (a);
      \draw (m) -- (ib) -- (b);
      \draw (m) -- (ic) -- (c);
      \draw (m) -- (id) -- (d);
      \draw (m) -- (ie) -- (e);
      \draw (m) -- (if) -- (f);
      \draw (m) -- (ig) -- (g);
      \draw (m) -- (ih) -- (h);
      \begin{pgfonlayer}{background}
        \node[draw,cloud,fit=(ia) (ib) (ic) (id) (ie) (if) (ig) (ih),
                   inner sep=-7pt,withshadow] (st) {};
        \node[fit=(d) (id),ellipse,inner sep=-1pt,rotate=20,draw,dashed,red] (sys) {};
      \end{pgfonlayer}
      \end{tikzpicture}\vspace*{-.5em}
  \caption{Interface theories}\label{fig:interface-theories}\vspace*{-1em}
\end{wrapfigure}
Before we go into any detail about how this ontology looks and how it induces a uniform
meaning space, we have to address another problem: the descriptions in the MitM ontology
must at the same time be system-near to make interfacing easy for systems, and serve as
an interoperability standard -- \emph{i.e.}\ be general and stable. If we have an ontology system
that allows modular/structured ontologies, we can solve this apparent dilemma by
introducing \defemph{interface theories}~\cite{KohRabSac:fvip11}, \emph{i.e.}\ ontology modules
(the light purple circles in Figure~\ref{fig:interface-theories}) that are at the same
time system-specific in their description of COMs -- near the actual representation of the
system and part of the greater MitM ontology (depicted by the cloud in
Figure~\ref{fig:interface-theories}) as they are connected to the core MitM ontology (the
blue circle) by views we call \defemph{interviews} (see below). The MitM approach
stipulates that interface theories and interviews are maintained and released together with
the respective systems, whereas the core MitM ontology represents the mathematical scope
of the VRE and is maintained with it. In fact in many ways, the core MitM ontology is the
conceptual essence of the mathematical VRE.

\subsection{Realizing and Utilizing a MitM Ontology}

\begin{wrapfigure}r{6.2cm}\centering\vspace*{-2em}
  \providecommand\myyscale{1}
\providecommand\myxscale{.9}
\begin{tikzpicture}[xscale=\myxscale,yscale=\myyscale]
\node[thy] (lf) at (0,2.5)  {$\cn{LF}$};
\node[thy] (lfx) at (1.8,2.5)  {$\cn{LF+X}$};
\node[thy] (fol) at (-1,1.5)   {$\cn{FOL}$};
\node[thy] (hol) at (.9,1.5) {$\cn{HOL}$};
\node[thy] (mon) at (-2.5,0) {$\cn{Monoid}$};
\node[thy] (gp) at (-.5,0) {$\cn{CGroup}$};
\node[thy] (rg) at (2,0)  {$\cn{Ring}$};
\node[thy] (zfc) at (-2.8,1.5) {$\cn{ZFC}$};

\draw[meta](lf) -- (fol);
\draw[meta](lf) -- (hol);
\draw[meta](fol) -- (mon);
\draw[meta](fol) -- (gp);
\draw[meta](hol) -- (rg);
\draw[include](lf) -- (lfx);
\draw[view](fol) -- node[above] {\footnotesize$\cn{f2h}$} (hol);
\draw[struct](gp) to[bend right=10] node[above] {\footnotesize$\cn{add}$} (rg);
\draw[struct](mon) to[out=20,in=160] node[above] {\footnotesize$\cn{mult}$} (rg);
\draw[include](mon) -- (gp);
\draw[view] (fol) -- node[above]{\footnotesize$\cn{folsem}$} (zfc);
\draw[view] (mon) -- node[right,near end]{\footnotesize$\cn{mod}$} (zfc);
\end{tikzpicture}\vspace*{-.5em}
  \caption{A OMDoc/MMT Theory Graph}\label{fig:mmt}\vspace*{-1em}
  \label{figure.omdoc.example}
\end{wrapfigure}
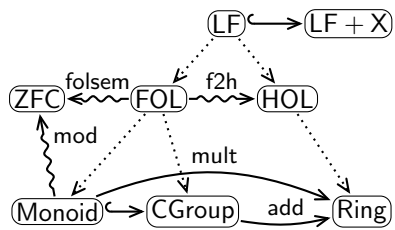
Our current candidate for representing the MitM ontology is the OMDoc/MMT
format~\cite{Kohlhase:OMDoc1.2,MMTSVN:on}. OMDoc/MMT is an ontology format specialized to
representing mathematical knowledge modularly in a theory graph: \defemph{theories} are
collections of declarations of concepts, objects, and their properties that are connected
by truth-preserving mappings called \defemph{theory morphisms}. The latter come in two
forms: \defemph{inclusions} and \defemph{structures} that essentially correspond to
object-oriented inheritance, and \defemph{view} that connect pre-existing theories -- in
these all axioms of the source theory have be to proven in the target theory. See
~\cite{RabKoh:WSMSML13} for a full account. Figure~\ref{fig:mmt} shows an example of
theory graph. It has three layers:
\begin{compactenum}[\em i\rm)]
\item the (bottom) \defemph{domain level}, which specifies mathematical domains as theories; here
  parts of elementary algebra. The hooked arrows are inclusions for inheritance, while the
  regular arrows are named structures that induce the additive and multiplicative
  structures of a ring.
\item the \defemph{logic level} represents the languages we use for talking about the
  properties of the objects at the domain level -- again as theories: the meta-theories of
  the domain-level ones -- the dotted arrows signify the meta-relation. At this level, we
  also have inclusions and views (the squiggly arrows) which correspond to logic
  translations (\cn{f2h}) and interpretations into \defemph{foundational theories} like
  set theory (here \cn{ZFC}). Incidentally models can be represented as views into
  foundations.
\item The top layer contains theories that act as metalogics, \emph{e.g.}\ the Logical Framework
  \cn{LF} and extensions which can be used to specify logics and their translations.
\end{compactenum}
The theory graph structure is very well-suited to represent heterogeneous collections of
mathematical knowledge, because views at the domain level can be used to connect differing
but equivalent conceptualizations and views at the logic level can be used to bridge the
different foundations of the various systems. The top level is only indirectly used in in
the MitM framework: it induces the joint meaning space via the meta-relation.

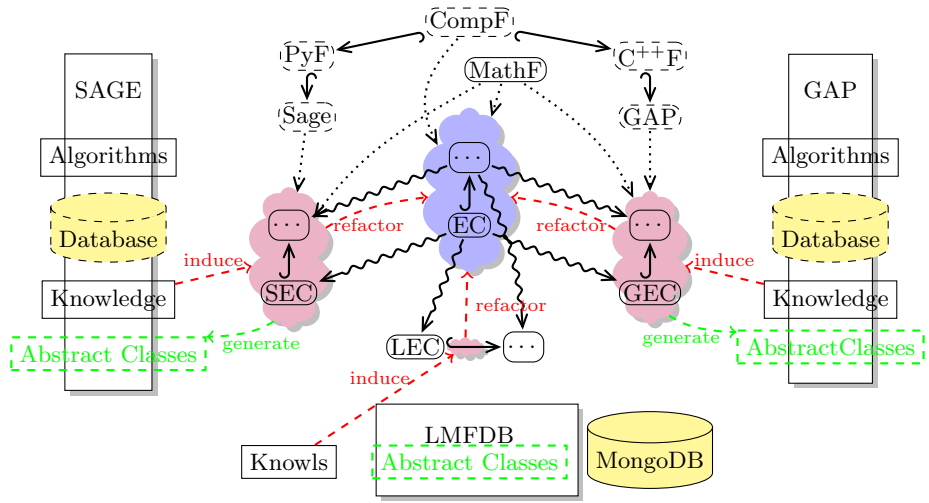
\begin{figure}[ht]\centering
  \begin{tikzpicture}[xscale=2.4,yscale=.9]
  \tikzstyle{withshadow}=[draw,drop shadow={opacity=.5},fill=white]
   \tikzstyle{database} = [cylinder,cylinder uses custom fill,
      cylinder body fill=yellow!50,cylinder end fill=yellow!50,
      shape border rotate=90,
      aspect=0.25,draw]
   \tikzstyle{human} = [red,dashed,thick]
   \tikzstyle{machine} = [green,dashed,thick]

\node[thy]  (mf) at (.2,5.3) {MathF};
\node[thy,dashed]  (compf) at (0,6) {CompF};
\node[thy,dashed]  (pf) at (-.9,5.5) {PyF};
\node[thy,dashed]  (cf) at (1,5.5) {C\textsuperscript{++}F};
\node[thy,dashed]  (sf) at (-0.9,4.6) {Sage};
\node[thy,dashed]  (gf) at (1,4.6) {GAP};

\draw[include] (compf) -- (pf);
\draw[includeleft] (compf) -- (cf);
\draw[include] (pf) -- (sf);
\draw[includeleft] (cf) -- (gf);

\node[thy] (kec) at (0,3) {EC};
\node[thy,minimum height=.4cm] (kl) at (0,4) {\ldots};

\node[thy] (sec) at (-1,2) {SEC};
\node[thy,minimum height=.4cm] (sl) at (-1,3) {\ldots};

\node[thy] (gec) at (1,2) {GEC};
\node[thy,minimum height=.4cm] (gl) at (1,3) {\ldots};

\node[thy] (lec) at (-.3,1.2) {LEC};
\node[thy,minimum height=.4cm] (ll) at (.3,1.2) {\ldots};

\node (sc) at (-2,5) {SAGE};
\node[draw] (salg) at (-2,4) {Algorithms};
\node[database,dashed] (sdb) at (-2,2.8) {Database};
\node[draw] (skr) at (-2,1.9) {Knowledge};
\node[draw,machine] (sac) at (-2,1.1) {Abstract Classes};

\node (gc) at (2,5) {GAP};
\node[draw] (galg) at (2,4) {Algorithms};
\node[database,dashed] (gdb) at (2,2.8) {Database};
\node[draw] (gkr) at (2,1.9) {Knowledge};
\node[draw,machine] (gac) at (2,1.2) {AbstractClasses};

\node (lmfdb) at (0,-.1) {LMFDB};
\node[database] (ldb) at (1,-.5) {MongoDB};
\node[draw] (knowls) at (-1,-.5) {Knowls};
\node[draw,machine] (lac) at (0,-.5) {Abstract Classes};

  \begin{pgfonlayer}{background}
    \node[draw,cloud,fit=(sec) (sl),aspect=.4,inner sep=-3pt,withshadow,purple!30] (st) {};
    \node[draw,cloud,fit=(gec) (gl),aspect=.4,inner sep=-4pt,withshadow,purple!30] (gt) {};
    \node[draw,cloud,fit=(kec) (kl),aspect=.4,inner sep=0pt,withshadow,blue!30] (kt) {};
    \node[draw,cloud,fit=(lec) (ll),aspect=3.5,inner sep=-8pt,withshadow,purple!30] (lt) {};
  \end{pgfonlayer}

\begin{pgfonlayer}{background}
  \node[draw,withshadow,fit=(sc) (skr) (sac) (sdb),inner sep=1pt] {};
  \node[draw,withshadow,fit=(gc) (gkr) (gac) (gdb),inner sep=1pt] {};
  \node[draw,withshadow,fit=(lmfdb) (lac) (ldb) (knowls),inner sep=1pt] {};
\end{pgfonlayer}

\draw[view] (kec) -- (sec);
\draw[view] (kec) -- (gec);
\draw[view] (kec) -- (lec);
\draw[include] (kec) -- (kl);
\draw[include] (gec) -- (gl);
\draw[include] (sec) -- (sl);
\draw[include] (lec) -- (ll);
\draw[view] (kl) -- (sl);
\draw[view] (kl) -- (gl);
\draw[view] (kl) to[bend left=5] (ll);

\draw[meta] (mf)  to [bend right=10] (st);
\draw[meta] (sf) -- (st);
\draw[meta] (mf)  to [bend left=10] (gt);
\draw[meta] (gf) -- (gt);
\draw[meta] (mf) -- (kt);
\draw[meta] (compf) to[bend right=15] (kt);

\draw[human,->] (skr) -- node[above]{\scriptsize induce} (st);
\draw[human,->] (gkr) -- node[above]{\scriptsize induce} (gt);
\draw[human,->] (knowls) -- node[left,near end]{\scriptsize induce} (lt);

\draw[machine,->] (gt) to[bend right=30] node[below,near start]{\scriptsize generate} (gac);
\draw[machine,->] (st) to[bend left=30] node[below,near start]{\scriptsize generate} (sac);
\draw[human,->] (st) to[bend left=20] node[below]{\scriptsize refactor} (kt);
\draw[human,->] (gt) to[bend right=20] node[below]{\scriptsize refactor} (kt);
\draw[human,->] (lt) -- node[right]{\scriptsize refactor} (kt);
\end{tikzpicture}
  \caption{The MitM Paradigm in Details}\label{fig:mitm}
\end{figure}
If we apply OMDoc/MMT to the MitM architecture, we arrive at the situation in
Figure~\ref{fig:mitm}, where we drill into the MitM information architecture from
Figure~\ref{fig:interface-theories}, but restrict at this stage to three systems from the \ODK
project. In the middle we see the core MitM ontology (the blue cloud) as an OMDoc/MMT
theory graph connected to the interface theories (the purple clouds) via MitM
interviews. Conceptually, the systems in \ODK consist of three main components:
\begin{compactenum}[\em i\rm)]
\item a \emph{Knowledge Representation component} that provides data structures for the
  COMs and their properties.
\item a \emph{DataBase component} that provides mass storage for objects, and 
\item a \emph{library of algorithms} that operate on these.
\end{compactenum}
To connect a system to an MitM-based VRE, the knowledge representation component is either
refactored so that it can generate interface theories, or a schema-like description of the
underlying data structures is created manually from which abstract data structures for the
system can be generated automatically -- in this version the interface theories act as an
Interface Description Language.

In this situation there are two ways to arrive at a greater MitM ontology: the \ODK
project aims to explore both: either
\begin{inparaenum}[\em i\rm)] 
\item standardizing a core MitM by refactoring the various interface theories where they
  overlap, or
\item flexiformalizing the available literature for a core MitM ontology.
\end{inparaenum}
For \emph{i}), the MitM interviews emerge as refinements that add system-specific details
to the general mathematical concepts\footnote{We use the word ``interface theory'' with a
  slightly different intention when compared to the original use
  in~\cite{KohRabSac:fvip11}: There the core MitM ontology would be an interface between
  the more specific implementations in the systems, whereas here we use the ``interface
  theories'' as interfaces between systems and the core MitM ontology. Technically the
  same issues apply.} For \emph{ii}), we have to give the interviews directly. 

To see that this architecture indeed gives us a uniform meaning space, we observe that the
core MitM ontology uses a mathematical foundation (presumably some form of set theory),
whereas the interface theories also use system-specific foundations that describe aspects
of the computational primitives of the respective systems. We have good formalizations of
the mathematical foundations already; first steps towards a computational ones have been
taken in~\cite{KohManRab:aumftg13}.

Our efforts also fit neatly alongside similar efforts underway across the sciences to
standardize metadata formats (for instance through the Research Data Alliance's Typing
Registry Working Group\cite{rda-typing}), except that the typing taking place here tends
to have much higher complexity since our objects of study are sometimes seen as types and
sometimes as instances (think of groups for instance).

\section{LMFDB Knowledge and Interoperability}\label{sec:lmfdb}
The \emph{$L$-functions and modular forms database} is a project involving dozens of
mathematicians, who assemble computational data about $L$-functions, modular forms, and
related number theoretic objects. The main output of the project is a website, hosted at
\url{http://www.lmfdb.org}, that presents this data in a way that could serve as a
reference for research efforts and should be accessible at the graduate student level.
The mathematical concepts underlying the \LMFDB are extremely complex and varied, so part
of the effort has been focused on how to relay knowledge (mathematical definitions and their
relationships) to data and software. For this purpose, the \LMFDB has developed so-called
\emph{knowls}, which are a technical solution to present \LaTeX-encoded information
interactively, heavily exploiting the concept of transclusion. The end result is a
very modular and highly interlinked set of definitions in mathematical vernacular.

The \LMFDB code is primarily written in \Python, with some reliance on \Sage for
the business logic. The frontend is written in the web framework Flask, while the backend
uses the NoSQL document database system \Mongo \cite{lmfdb-repo}. Again, due to the
complexity of the objects considered, many idiosyncratic encodings are used for the
data. This makes the whole data management lifecycle particularly tricky, and dependent on
different select groups of individuals for each component.

As the LMFDB spans the whole ``vertical'' workflow, from writing software, to producing new
data, up to presenting this new knowledge, it is a perfect test case for a large scale
case study of the MitM approach. Conversely, a semantic layer would be beneficial to its
activities across data, knowledge and software, which it would help integrate more
cohesively and systematically.

Among the components of the LMFDB, elliptic curves stand
out in the best shape, and a source of best practices for other areas. 
We have generated MitM interface theories for LMFDB elliptic curves by (manually)
refactoring and flexiformalizing the {\LaTeX} source of knowls into \sTeX (see
Listing~\ref{stex-ec} for an excerpt), which can be converted into flexiformal OMDoc/MMT
automatically. The MMT system can already type-check the definitions, avoiding circularity
and ensuring some level of consistency in their scope and make it browsable through
\textsf{MathHub.info}, a project developed in parallel to MMT to host such formalisations.

\begin{lstlisting}[language={[sTeX]TeX},label={stex-ec},firstline=31,
   caption= {\protect\stex flexiformalization of an \LMFDB knowl}]
\begin{mhmodnl}{minimal-Weierstrass-model}{en}
    A \defi{minimal} \trefii{Weierstrass}{model} is one for which
    $\absolutevalue\wediscriminantOp$ is minimal among all Weierstrass models
    for the same curve.  For elliptic curves over $\RationalNumbers$, minimal 
    models exist, and there is a unique minimal model which satisfies the 
    additional constraints $\minset{\livar{a}1,\livar{a}3}{\set{0,1}}$, and 
    $\inset{\livar{a}2}{\set{-1,0,1}}$. 
    This is defined as the reduced minimal Weierstrass model of the elliptic curve.
  \end{definition}
\end{mhmodnl}
\end{lstlisting}

   The second step consisted of translating these informal definitions into progressively
   more exhaustive MMT formalisations of mathematical concepts (see
   Listing~\ref{lst:mmt-ec}). The two representations are coordinated via the theory and
   symbol names -- we can see the \sTeX representation as a human-oriented documentation
   of the MMT.

\begin{lstlisting}[morekeywords={namespace,theory,include},mathescape,
firstline=21,
caption= {MMT formalisation of elliptic curves and their Weierstrass models},
label=lst:mmt-ec]
theory minimal_Weierstrass_model : odk:?Math =
  include ?elliptic_curve $\RS$
  minimal : tm Ws_model  $\rightarrow$  tm Ws_model $\RS$
  is_minimal : tm Ws_model  $\rightarrow$  prop \US = [A] (minimal A) $\doteq$ A $\RS$
  minimality_idempotence : {A} $\vdash$ minimal (minimal A) $\doteq$ minimal A $\RS$
  minimality_of_minimal_Ws_model : 
    {A} $\vdash$ is_minimal (minimal_Ws_model A)  $\RS$
  injective_minimal_Ws_model : 
    {A,B} $\vdash$ minimal_Ws_model A $\doteq$ minimal_Ws_model B  $\rightarrow$ $\vdash$ A $\doteq$ B  $\RS$
$\GS$
\end{lstlisting}

Finally, we have to integrate computational data into the interface theories. Based on
recent ongoing efforts \cite{lmfdb-formats} to document the \LMFDB ``data schemata'' we
established OMDoc/MMT theories that linked the database fields to their data types (string
\emph{vs.} float \emph{vs.} integer tuple, for instance) and mathematical types (elliptic
curves or polynomials) -- the latter based on the vocabulary in the interface theories
generated from the \LMFDB knowls. This schema theory is complemented by a theory on
composable \emph{MMT codecs}, which in turn acts as a specification for a collection of
implementations in various programming languages (currently \Python, Scala, and
C\textsuperscript{++} for \Sage, MMT, and \GAP respectively) which are first instances of
a computational foundation (see Section~\ref{sec:mitm}).  For instance, one could compose
two MMT codecs, say \emph{polynomial-as-reversed-list} and
\emph{rational-as-tuple-of-int}, to signify that the data $[(2,3),(0,1),(4,1)]$ is meant
to represent the polynomial $4x^2+2/3$. Of course, these codecs could be further
decomposed (signalling which variable name to use, for instance). The initial cost of
developing these codecs is high, but the clarity gained in documentation is valuable, they
are highly reusable, and they drastically expand the range of tooling that can be built
around data management.

\paragraph{A typical application}
Based on these MitM interface theories we can generate I/O interfaces that translate
between the low-level \LMFDB API, which delivers raw \Mongo data in JSON format into MMT
expressions that are grounded in the interface theories. This ties the \LMFDB database
into the MitM architecture transparently. As a side effect, this opens up the \LMFDB to
programmatic queries via the MMT API, which can be queried and can then relay them to the
\LMFDB API directly and transparently.

\section{Distributed Collaboration with GAP/Sage}\label{sec:gapsage}
\label{sec:handles}

Another aspect of interoperability in a mathematical VRE is the possibility of distributed
multisystem computations, where \emph{e.g.}\ a given system may decide to delegate
certain subcomputations or reasoning tasks to other systems.

There are already a variety of peer-to-peer interfaces (see Figure~\ref{fig:interop})
based on the ``handle paradigm'' between systems in the \ODK project;
for example \Sage includes interfaces for \GAP, \Singular, or \Pari.

In the ``handle paradigm'', when a system $A$ delegates a calculation to a system $B$, the
result $r$ of the calculation is not converted to a native $A$ object; instead $B$ just
returns a handle (or reference) to the object $r$. Later $A$ can run further calculations
with $r$ by passing it as argument to $B$ functions or methods. The advantages of this
approach include that we can avoid the overhead of back and forth conversions between $A$
and $B$ and that we can manipulate objects of $B$ from $A$ even if they have no native
representation in $A$.

Given a mapping of corresponding methods in the systems, we can use the adaptor pattern to
implement this. For example, calling the method \texttt{h.cardinality()} on a \Sage handle
\texttt{h} to a \GAP object \texttt{G}, triggers in \GAP a call to \texttt{Size(G)} if
\texttt{cardinality} and \texttt{Size} are marked as corresponding. But this dispatch
depends on an alignment of the type systems in \Sage and \GAP. For example, if \texttt{h}
is a handle to a set \texttt{S}, \Sage only knows that \texttt{h.cardinality()} can be
computed by \texttt{Size(S)} in \GAP if \texttt{S} is a group; in fact if \texttt{h} has
been constructed through the \texttt{PermutationGroup} or \texttt{MatrixGroup}
constructors. Whereas we would want this method to be available as soon as \texttt{S} is a
set.

To get around this problem we have worked on a more semantic integration, where adaptor
methods are made aware of the type hierarchies of the respective other system, see
Listing~\ref{lst:adaptor} below. 
\begin{lstlisting}[language=Python,label=lst:adaptor,
  caption=A Semantic Adaptor Method in \Sage]
class Sets: # Everything generic about sets in Sage
    class GAP: # The adapter methods relevant to Sets in the Sage-Gap interface
         class ParentMethods: # Adapter methods for sets
             def cardinality(self): # The adapter for the cardinality method
                 return self.gap().Size().sage()
         class ElementMethods: # Adapter methods for set elements
             ...
         class MorphismMethods: # Adapter methods for set morphisms
             ...
\end{lstlisting}

This peer-to-peer approach however does not scale up to a dozen of
systems. This is where the MitM paradigm comes to the rescue. With it,
the task is reduced to building interface theories and interviews into
the core MitM ontology, in such a way that the adaptor pattern can be
made generic in terms of the MitM ontology structure, without relying
on the concrete structure of the respective type systems. Then the
adapter methods for each peer-to-peer interface can be automatically
generated.

In our example, the correspondence between \texttt{cardinality} and
\texttt{Size} still holds if the MitM interviews link the
\texttt{cardinality} function in the \Sage interface theory on sets
with the \texttt{Size} function in the corresponding interface theory
for \GAP.

We will now show first results of our experiments with interface
theories and interviews, including several applications beyond the
generation of interface theories that support distributed computation
for \Sage and \GAP.

\subsection{Semantics in the \Sage Category System}

The \Sage library includes 40k functions and allows for manipulating
thousands of different kinds of objects. As usual in such large
systems, it is critical for taming code bloat to
\begin{compactenum}[\em i\rm)]
\item identify the core concepts describing common behavior among the objects;
\item exploit this to implement generic operations that apply on all object having a given
  behavior, with appropriate specializations when performance calls for it;
\item design or choose a process for selecting the best implementation available when
  calling an operation on one or several objects.
\end{compactenum}

Following mathematical tradition and the precedent of the \Axiom,
\Fricas, or \MuPAD systems, \Sage has developed a
category-theory-inspired ``category system'', and found a way to
implement it on top of the underlying \Python object
system~\cite{Sage,Sage.Categories}. In short, a \defemph{category} specifies
the available \defemph{operations} and the \defemph{axioms} they satisfy.
This category system models taxonomic knowledge from mathematics
explicitly and uses it to support genericity, control the method
selection process, structure the code and documentation, enforce
consistency, and provide generic tests.

\begin{wrapfigure}r{8cm}\vspace*{-2.5em}
\begin{lstlisting}[language=Python]
@semantic(mmt="sets")
class Sets:
    class ParentMethods:
         @semantic(mmt="card?card", gap="Size")
         @abstractmethod
         def cardinality(self):
             "Return the cardinality of ``self``"
\end{lstlisting}
\vspace*{-.5em}
\caption{An annotated category in \Sage}\label{fig:anncat}\vspace*{-1.5em}
\end{wrapfigure}
To generate interface theories from the \Sage category system, we are experimenting with a
system of annotations in the \Sage source files. Consider for instance the situtation in
Figure~\ref{fig:anncat} where we have annotated the \texttt{Sets()} category in \Sage
with \texttt{@semantic} lines that state correspondences to other interface theories. From
these the \Sage-to-MMT exporter can generate the respective interface theories and views.

Several variants of the annotations are experimented with to allow for adding annotations on existing
categories without touching their source file, and also for specifying directly the corresponding
method names in other systems when this has not yet been formalized elsewhere. Similarly,
one can provide directly the signature information in case that is not yet modelled in
MMT.

\subsection{Exporting the \GAP Knowledge: Type System Documentation}
\label{sec:gaptypes}

As in \Sage, the \GAP type system encodes a wealth of mathematical
knowledge, which can influence method selection. For example
establishing that a group is nilpotent will allow for more efficient
methods to be run for finding its centre. The main difference lies in
the method selection process. In \Sage the operations
implemented for an object and the axioms they satisfy are specified by
its class which, together with its super classes, groups syntactically
all the methods applicable in this context. In \GAP, this information
is instead specified by the truth-values of a collection of
independent \defemph{filters}, while the context of applicability is
specified independently for each method.
Breuer and Linton describe the \GAP type system in \cite{breuer-linton} and
the \GAP documentation \cite{GAP4} also contains extensive information on the types
themselves.

\begin{wrapfigure}r{6cm}\vspace*{-2em}
  \includegraphics[width=6cm]{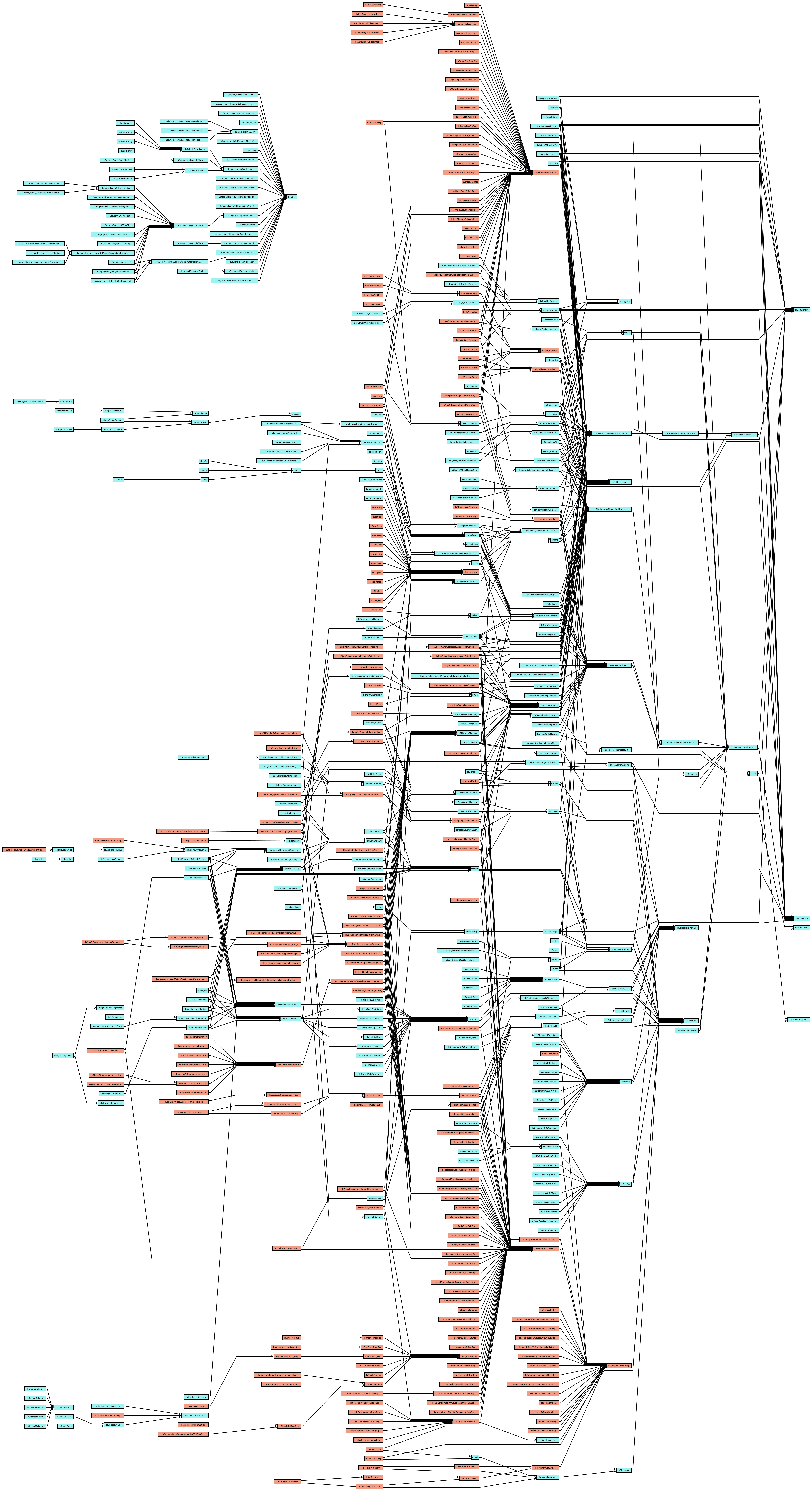}\vspace*{-.5em}
  \caption{The \GAP Knowledge Graph.\label{fig:gap-graph}}\vspace*{-2em}
\end{wrapfigure}
\GAP allows some introspection of this knowledge after the system is loaded: the values of
attributes and properties can be unknown on creation, can be computed on demand, and their
values can then be stored for later reuse without the need to be recomputed.

\begin{figure}[ht]\centering
  \includegraphics[width=\textwidth]{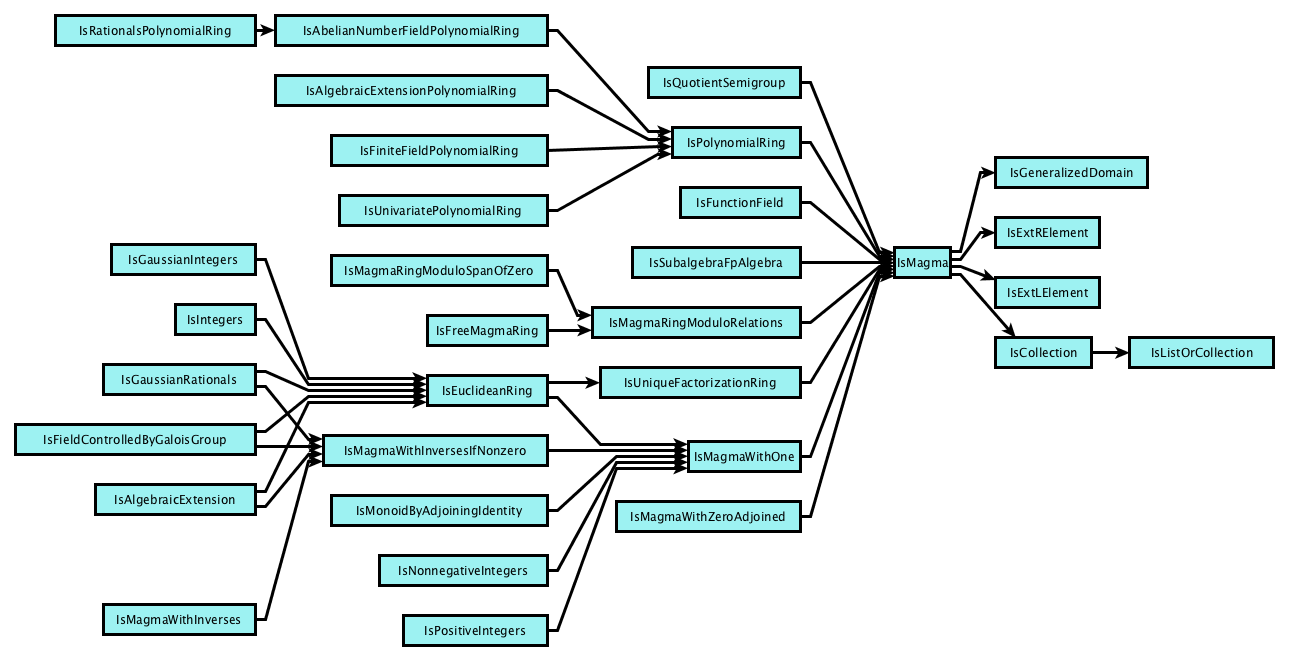}
  \caption{The \GAP Knowledge Graph (fragment).\label{fig:gap-ismagma}}
\end{figure}

As a first step in generating interface theories for the MitM ontology, we have developed
tools to accesses mathematical knowledge encoded in \GAP, such as introspection inside a
running \GAP session, export to JSON to import to MMT, and export as a graph for
visualisation and exploration. These will become generally available in the next \GAP
release. The JSON output of the \GAP object system with default packages is currently
around 11 Megabytes and represents a knowledge graph with 540 vertices, 759 edges and 8 connected
components, (see Figures~\ref{fig:gap-graph},\ref{fig:gap-ismagma}). If all
packages are loaded, this graph expands to 1616 vertices, 2178 edges and 17 connected
components.

There is however another source of knowledge in the \GAP universe: the documentation, which is
provided in the special format GAPDoc \cite{gapdoc}. Besides the main manuals the GAPDoc
format is adopted by 97 out of 130 packages currently redistributed with
GAP. Conventionally GAPDoc is used to build text, PDF and HTML versions of the manual
from a common source given in XML. The GAP reference manual is almost 1400 pages and the
packages add hundreds more.

The GAPDoc sources classify documentation by the type of the documented object (function,
operation, attribute, property, etc.) and index them by system name. In this sense they
are synchronized with the type system (which \emph{e.g.}\ has the types of the functions) and can
be combined into flexiformal OMDoc/MMT interface theories, just like the ones for \LMFDB
in Section~\ref{sec:lmfdb}. This conversion is currently under development and will lead
to a significant increase of the scope of the MitM ontology. 

As a side-effect of this work, we discovered quite a few inconsistencies in the \GAP
documentation which came from a semi-automated conversion of GAP manuals from the
\TeX-based manuals used in GAP 4.4.12 and earlier.  We developed the consistency checker
for the GAP documentation, which extracts type annotations from the documented GAP objects
and compares them with their actual types. It immediately reported almost 400
inconsistencies out of 3674 manual entries. In the subsequent cleanup, we by now have
eliminated about 75\% of them.

\section{Conclusion}\label{sec:concl}
In this paper we have presented the \ODK project and the ``Math-in-the-Middle'' approach
it explores for mitigating the system integration problems inherent in combining an ecosystem
of open source software systems into a coherent mathematical virtual research environment.
The MitM approach relies on a central, curated, flexiformal ontology of the mathematical
domains to be covered by the VRE together with system-near interface theories and
interviews to the core ontology that liaise with the respective systems. We have reported
on two case studies that were used to evaluate the approach: an interface for the \LMFDB,
and a more semantic handle interface between \GAP and \Sage.

Even though the development of the MitM is still at a formative stage, these case studies
show the potential of the approach. We hope that the nontrivial cost of curating an
ontology of mathematical knowledge and interviews to the interface theories will be offset
by its utility as a resource, which we are currently exploring; the unification of the
knowledge representation components
\begin{compactitem}
\item enables VRE-wide domain-centered (rather than system-centered) documentation;
\item can be leveraged for distributed computation via uniform protocols like the
  SCSCP~\cite{HorRoz:ossp09} and MONET-style service
  matching~\cite{CaprottiEtAl:MathServiceMatching04:tr} (the absence of content
  dictionaries -- MitM theories -- was the main hurdle that kept these from gaining more
  traction);
\item will lead to the wider adoption of best practices in
  mathematical knowledge management in the systems involved -- in
  fact, this is already happening.
\end{compactitem}
Whether in the end the investment into the MitM will pay off also depends on the quality
and usability of the tools for mathematical knowledge management. Therefore we invite the
CICM community to interact with and contribute to the \ODK project, on
this work package and the others.

\subsubsection*{Acknowledgements}

The authors gratefully acknowledge the other participants of the St~Andrews workshop, in
particular John Cremona, Luca de Feo, Mihnea Iancu, Steve Linton, Dennis M\"uller, Viviane
Pons, Florian Rabe, and Tom Wiesing, for discussions and experimentation which clarified
the ideas behind the math-in-the-middle approach.

We acknowledge financial support from the OpenDreamKit Horizon 2020 European Research
Infrastructures project (\#676541), from the EPSRC Collaborative Computational Project
CoDiMa (EP/M022641/1) and from the Swiss National Science Foundation grant PP00P2\_138906.

\printbibliography
\end{document}